# Biaxial compression of centimeter scale graphene on strictly 2D substrate

Hadi Arjmandi-Tash[1], Hessam Sokooti[2], Khosrow Shakouri[1], Lin Jiang[1], Alexander Kloosterman[1], Lia M.C. Lima[1], and Grégory F. Schneider[1*]

[1] *Faculty of Science, Leiden Institute of Chemistry, Leiden University, Einsteinweg 55, 2333CC Leiden, The Netherlands*

[2] *Division of Image Processing, Leiden University Medical Centre, 2300 RC Leiden, The Netherlands*

[†]to whom correspondence should be addressed: g.f.schneider@chem.leidenuniv.nl

We study the mechanics of graphene floating on the surface of water, subject to biaxial compression in a Langmuir-Blodgett (LB) trough. The platform uniquely allows to characterize graphene samples with centimeter-scale lengths, four orders of magnitude larger than commonly probed microscale graphene samples. We plot full stress-strain diagram of graphene identifying elastic and plastic deformation regions. The Young's modulus respects a predicted scaling law and falls two orders of magnitude below the previous reports. The rigidity of graphene against out-of-plane deformations (flexural rigidity), on the other hand, strongly improves in our large samples. Our results demonstrate that graphene − in its very natural form − lacks any intrinsic elastic parameters. Functionalization/manipulation of graphene lattice affects the mechanics of graphene; particularly we explored the effect of the $sp^3$ hybridization and crystalline voids. Compression of graphene beyond the elastic limit yields wrinkles; evolution of the wrinkles in a progressive compression gives rise to internal friction, i.e. viscoelastic characteristics, recognized for the first time in graphene. Our results elucidate yet-unexplored science of graphene and establish novel applications for this material in mechanical systems.

Picturing graphene (and similar materials) as a strictly two-dimensional sheet is unrealistic; indeed though dubbed as two-dimensional, the surface of such materials exhibit transverse out-of-plane undulations (the bending modes) with limited amplitudes in addition to the expected in-plane undulations (stretching modes). Both the in-plane and transverse undulations contribute to the total elastic energy. Harmonic approximation provides a fast solution for this energy considering that the dual modes are decoupled. In this approximation, the average amplitude of the transverse undulations scales with $L^2$ (L: size of the sample) and approaches infinity for large samples (*1*). This result was interpreted as the tendency for crumpling which questioned the existence of graphene for several decades (*2*). A more realistic solution, however, was achieved considering that the bending and stretching modes are coupled and mutually affect each other (anharmonic approximation). The transverse displacements are now considerably weaker and scale with $L^{1-\eta/2}$ ($\eta$ = 0,8), though still larger than the interatomic distances. Indeed the approximation predicts an intrinsic tendency for small amplitude transverse undulations which eventually and thermodynamically stabilizes the material. The undulations are still size-dependent which activate size-dependent elastic properties. Particularly atomistic Monte Carlo (MC) simulations demonstrates that the out-of-plane undulations in anharmonic approximation renormalize the in-plane stiffness of graphene as ~ $L^{-0.325}$ (L: size of the sample) (*3*). The effect, however, cannot be experimentally measured using existing standard platforms (e.g. nano-indentation experiments (*4–8*) and graphene nano-resonators (*9–11*)) where graphene is subject to tensile loadings. Indeed the tensile strain suppresses the anharmonicity (*12*). The lack of a suitable experimental approach, on the other hand, has limited the knowledge about the performance of graphene under compression. In fact compressing free-standing graphene is impractical as graphene sags. Once supported on a rigid substrate, however, the out-of-plane deformations inherited from the corrugations on the surface of the substrate and graphene/substrate interactions affect the experiment (*13*, *14*). More importantly, the mechanics of the substrate is involved in the measurements (*15*) and decoupling the contributions of graphene and the substrate is complex. Such experiments may lack accuracy also as a result of the lateral slippage of graphene during the experiments (*16*, *17*).

Water offers numerous advantages to conduct graphene compression experiments, hardly available with rigid substrates: i) The surface of water is strictly planar and serves as an appropriate mattress for the two-dimensional graphene. ii) The adhesion energy between hydrophobic graphene and water is lower than graphene on typical substrates (*18*) (45 times lower than graphene on silicon) which guarantees a negligible graphene/substrate interaction. In addition, iii) graphene can smoothly slide on the surface of water (no dry friction) which allows free deformation upon compressive buckling. We note that iv) ultra-pure water is easily accessible; the uniform and simple chemistry of the water mattress is not a source of irreproducibility, demonstrating that it is now possible to probe still unexplored properties of graphene.

We performed the study on a LB trough, capable of applying a precisely controlled two dimensional pressure on the sample under test while freely floating at air/water interface (*19*). A piece of ~1 × 1 $cm^2$ crack-free (*20*) (Supplementary Section S3) graphene, chemically grown on a copper foil (*21*), was placed on the surface of a shallow ammonium persulfate solution (0.5 M) on the trough and surrounded with Dipalmitoylphosphatidylcholine (DPPC) lipid molecules (Supplementary Sections S1 and S2). Once the copper foil etches away, graphene turns to be freely floating at the air/liquid interface. Unlike the compression of graphene embedded in polymeric materials where the placement of graphene or curing the supporting polymer deposit a residual strain (*16*) (which adds complexity to the analysis), graphene in our experiments can relax to release any strain (e.g. due to the growth) before initiating the measurements. Moving two slightly-wetted solid barriers − initially positioned at opposite sides of the trough − forward (closer to each other) lowers the enclosed area which eventually applies a two-dimensional surface pressure *p* (we refer to as stress in this letter) to graphene (Figure 1-a and Supplementary Video). The role of the lipid molecules is pronounced: they form a stable, two-dimensional and dynamic self-assembled layer at the air/liquid interface where graphene lies and guarantee a strictly lateral compression (no perpendicular component) of graphene. The Wilhelmy plate of the LB trough records *p*. Additionally, we estimated the two-dimensional strain ($\varepsilon_{||} = {(A_0 - A)}/{A_0}$, $A_0$: initial area of graphene, $A$: area of graphene at the time of the measurement) of the graphene by processing high resolution images taken by a camera focusing on graphene (Supplementary Sections S1 and S2).

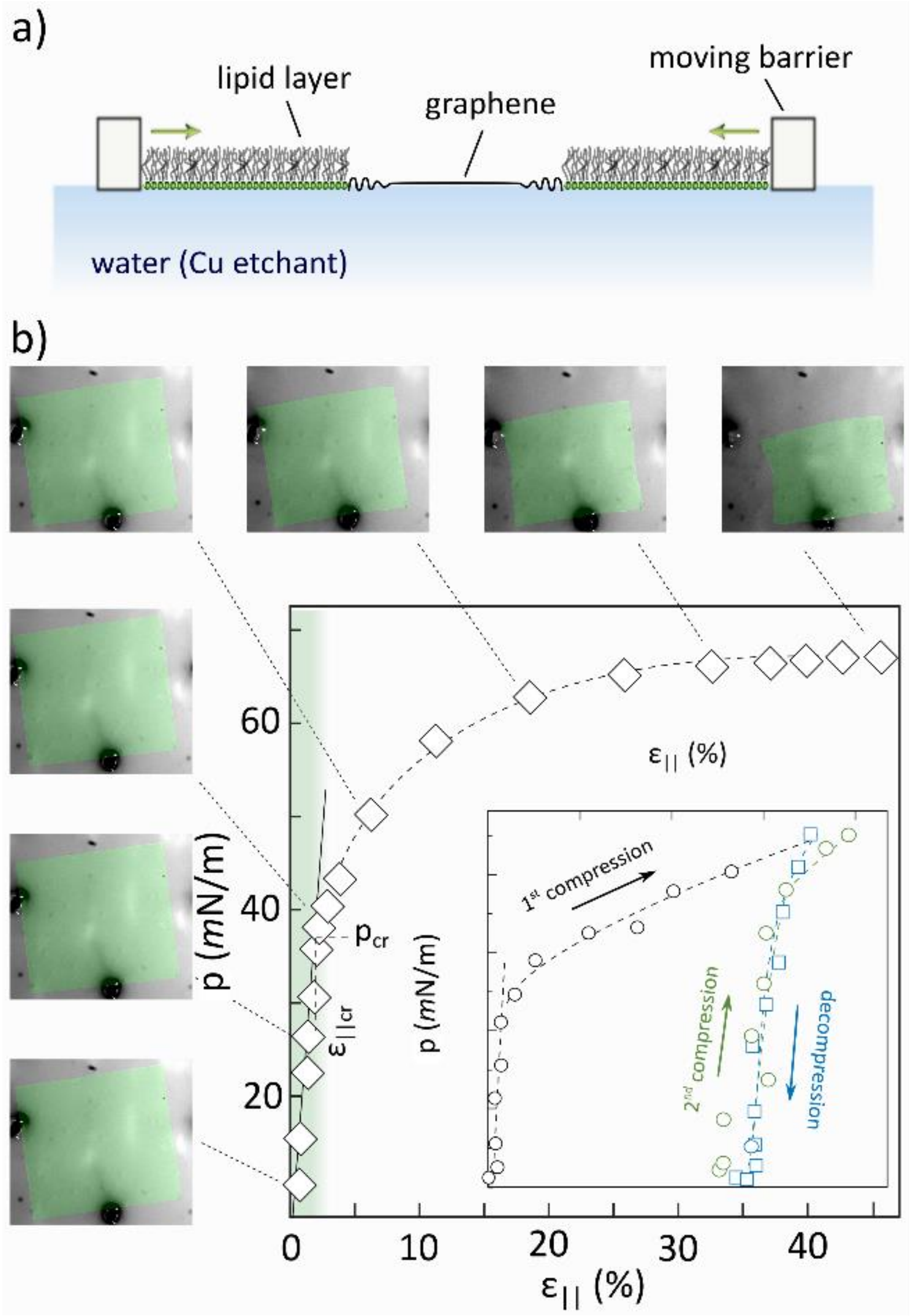

Figure 1: Floating graphene subject to 2D compression

a) Schematic representation of the experiment set-up: graphene floating on the surface of a copper etchant (0.5 M ammonium persulfate) is subject to a compressive stress applied by surrounding lipids and moving barriers.

b) Stress-strain relation of a floating graphene piece (~1 × 1 $cm^2$): The green shaded area in the main plot corresponds to the elastic deformation region with the solid line as the linear fit. Few snapshots of the graphene (colored in green) at different compression stages are presented at the left and top margins, featuring progressive deformations upon compression. Inset panel corresponds to the subsequent compression (black)/decompression (blue)/compression (green) of a separate graphene sheet. The horizontal displacement between the first and second compression strokes corresponds to the permanent plastic deformation. The barrier displacement rate was limited to 1 mm/min in these experiments.

Figure 1-b plots the stress-strain curve of a selected graphene sample. Similar to conventional materials, graphene exhibits an elastic deformation (linear stress-strain relation) up to a certain level (critical stress $p_{cr}$); Then plastic deformation, characterized by the buckling of graphene starts. Unlike elastic deformation

which is reversible, plastic deformation is permanent: removing the external pressure, does not return graphene to its original shape (inset Figure 1-b). Bulk modulus ($k$) in biaxial loading measures the resistance of the material against elastic deformation: $p = k\varepsilon_{||}$ (*22*) and scales with the two dimensional Young's modulus ($E_{2D}$) as: $k = {E_{2D}}/{2(1-\vartheta)}$ (*22*) where $\vartheta = 0.165$ is the Poisson's ratio of the graphene.

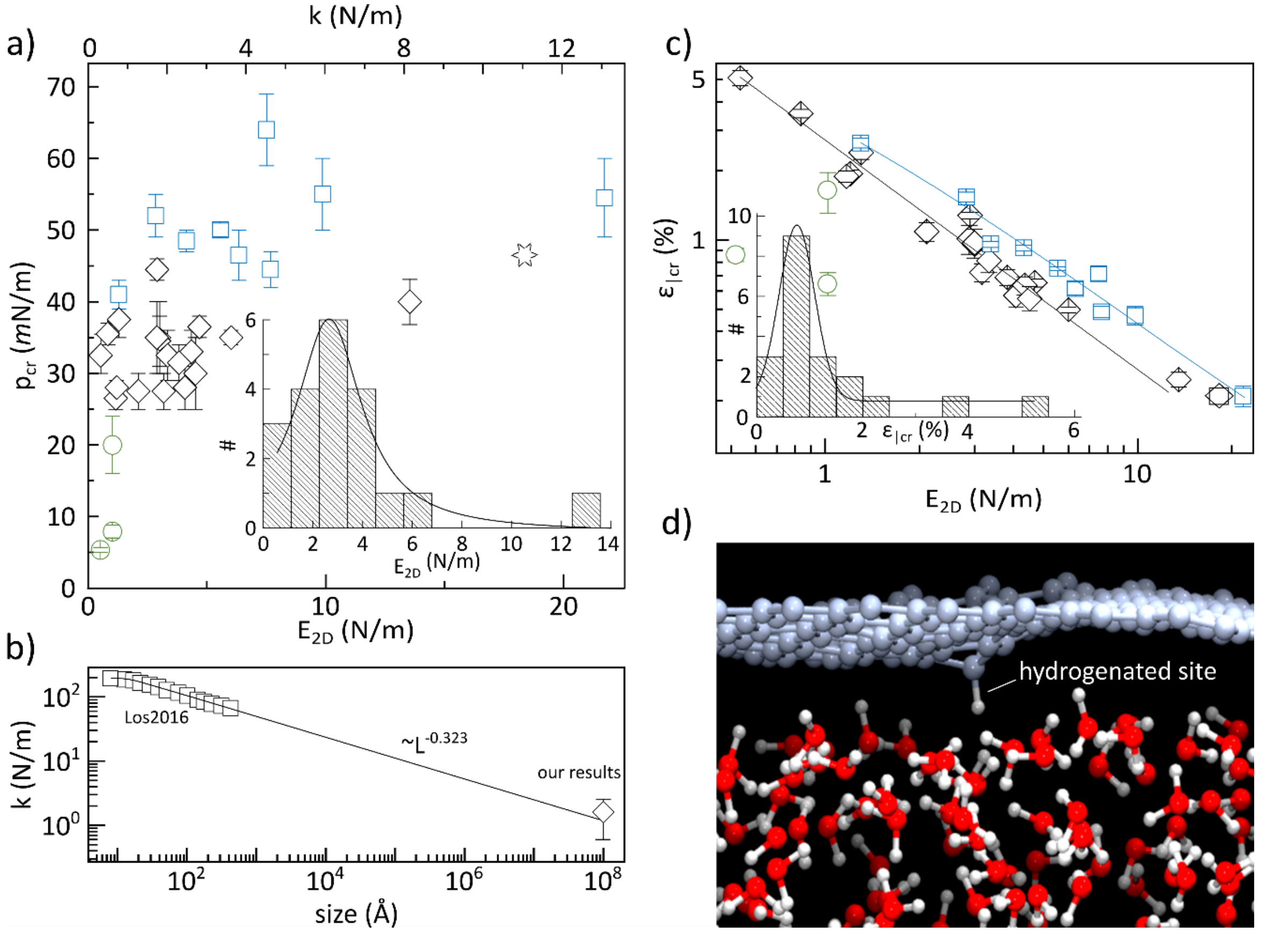

Figure 2: Elastic properties of graphene subject to 2D compression

a) Correlation of the yield strength and in-plane stiffness of untreated (monolayer: black diamonds, bilayer: black star), hydrogenated (blue squares), $Ar^+$ bombarded (green circles) graphene. The inset displays the distribution of the $E_{2D}$ of untreated monolayer graphene samples, centered at 2.6 N/m with the FWHM of 3.2 N/m estimated by Lorentzian fitting (solid line).

b) Superimposition our experimentally measured bulk moduli (untreated graphene) with the simulation results of nanoscale samples, by Los et. al. (*3*): the solid line corresponds to the scaling law reported by the same group(Supplementary Section S4).

c) Correlation of the critical one-dimensional strain $\varepsilon_{|cr}$ $(\approx \varepsilon_{||cr}/2)$ and in-plane stiffness of samples in (a) (the same marker and color codes apply here); the solid lines correspond to the best fittings with $\varepsilon_{|cr} = \frac{\alpha}{(E_{2D}/\beta)}$ revealing α = 0.027 N/m and β = 1 for untreated and α = 0.027 N/m and β = 1.4 for hydrogenated graphene. Inset shows the distribution of the $\varepsilon_{|cr}$ of untreated monolayer graphene, centred at 0.8%.

d) Molecular dynamics simulation of hydrogenated graphene on the surface of water: interaction between the hydrogenated site and water molecules locally stabilizes the corresponding carbon atom. The snapshot corresponds to $\varepsilon < \varepsilon_{cr}$ region.

It is commonly expressed that pristine graphene, possessing strong bonds between the carbon atoms, is one of the most stiff materials (few hundred times stiffer than steel (*4*)) featuring a Young's modulus of $E_{2D}^{flat} \sim 336\ {}^{N}/_{m}$ (*23–26*). Such a statement has to be used conservatively as its validity is limited to graphene samples that are freed from out-of-plane perturbations (flat graphene). Indeed tensile loading of the samples, as the standard procedure to probe the stiffness of graphene so far (*4–11*), inevitably flattens graphene and suppress anharmonicity (*12, 27*). Our approach is the only existing method so far which measures the stiffness of graphene while the natural perturbations remain untouched. The median Young's moduli of our twenty monolayer graphene samples approaches $\bar{E}_{2D}$ = 2.6 N/m (k = 1.6 N/m) (Figure 2-a). Remarkably, this value is two orders of magnitude smaller than $E_{2D}^{flat}$ but can be well-explained in the framework of the scaling effect in the presence of the anharmonicity (Figure 2-b and Supplementary Section S4): Indeed for large samples, the lower energy costs of out of plane fluctuations ($\propto q^2$, q: the wave vector) compared to in-plane ones ($\propto$ q) favors the long wavelength out-of-plane fluctuations and increases the contribution of the anharmonic energy correction terms in total elastic energy. The size dependency of the energy terms causes a size dependent stiffness which falls in large samples. Interestingly the stiffness of our centimeter-scale samples falls on the extrapolation and respects the same scaling law of the samples of few (tens of) nanometer size (~ 6 orders of magnitude smaller in size, ~12 orders of magnitude larger in area, more relevant for 2D compression) and validates the accuracy of those modelling.

Manipulation of the lattice affects the mechanics of graphene (Supplementary Sections S5 and S6). Supplementary Figure S4-a compares the stress-strain relation of graphene samples with vacancy defects of different concentrations (achieved by $Ar^+$ bombardment of the graphene lattice, Supplementary Section S1 for the Methods and S5 for the quantitative measurement of defect density). Inclusion of the vacancy defects dramatically lowers $p_{cr}$: Here the graphene does not exhibits the typical transformations of different wrinkle types upon increasing the external pressure, leading to a fast (catastrophic) failure (Supplementary Sections S9 and S10).

The inclusion of $sp^3$ defects (achieved by hydrogenation of the graphene lattice) considerably improves $p_{cr}$. Non-covalent graphene/substrate interaction has been observed to delay buckling by preventing the delamination of graphene from polymeric substrates (*13, 16*). In a similar scenario, one could attribute the increase of $p_{cr}$ to the apparition of hydrogen bonds between hydrogenated graphene and water in our system. In Supplementary Section S7 and S8, however, we theoretically illustrate that water conforms graphene during the buckling and no delamination is involved. In fact, the increment in $p_{cr}$ in hydrogenated graphene is a result of their improved Young's modulus, discussed below.

Figure 2-c plots the $\varepsilon_{|\mathrm{cr}}(\approx \varepsilon_{||cr}/2)$ of different samples as a function of the stiffness. Untreated and hydrogenated samples respect the classical Eulerian picture of buckling (*13*) and exhibit $\varepsilon_{|\mathrm{cr}} \propto {}^{1}/_{\mathrm{E}_{2\mathrm{D}}}$ behaviour. This trend is not observable with the $Ar^{+}$ bombarded samples, further confirming their distinct failure mode. Notably, the hydrogenation of graphene improves the Young's modulus by β = 1.5 times, matching with the improvement in $p_{cr}$ discussed in Figure 2-a. In the framework of the anharmonic coupling the stretching and flexural modes in 2D materials and similar to an earlier observation (*8*), we attribute the improvement to the partial screening and eventual localization of the long wavelength flexural modes by $sp^3$ defects in hydrogenated graphene (Figure 2-d). Particularly the improvement factor β is reasonably close to $\left({}^{n_d}/_{n_0}\right)^{\eta/2}$ (η = 0.36, $n_d$: density of the $sp^3$ defects ranging from $5 \times 10^{10}$ cm$^{-2}$ to $1 \times 10^{12}$ cm$^{-2}$, $n_0$: density of the defects in untreated graphene estimated as $9 \times 10^{9}$ cm$^{-2}$ by Raman spectroscopy, Supplementary Section S5) predicting 1.3 to 2.3 times improvement in Young's modulus (*8*). Our molecular dynamics simulations also confirms that the flexural modes are affected by hydrogenated sites, leading to a 1.35 times improvement in Young's modulus, matching well with the experimental measurement (Supplementary Figure S5-b and c).

In the classical Eulerian picture of buckling, the coefficient α (see the fitting in Figure 2-c) depends on the flexural rigidity *D* of the specimen. In Supplementary Section S11 and by analyzing the buckling patterns we estimate *D* to be in the order of $10^{-12}$ Nm. The value for our centimeter-scale samples is at least three orders of magnitude larger than the previous report for micro-sized graphene samples (*28*) which itself is three to four orders of magnitude larger than for flat graphene. The trend agrees with the predicted scaling properties of flexible membranes (*29*). Our observations of scaling both the Young's modulus and flexural rigidity demonstrates that graphene, in its natural form, lacks any intrinsic elastic coefficients.

Inset in Figure 2-c plots the distribution of the critical strain of untreated graphene samples. Interestingly, the median critical strain ($\bar{\varepsilon}_{|\mathrm{cr}}$ = 0.8%) in our centimetre-scale samples (L/t ~ $10^{8}$) approaches that of narrow (micro-sized) graphene samples sandwiched in between polymeric supports (*13*); exhibiting such a large strain before buckling is attributed to the strictly 2D surface of water which excludes any substrate-induced out-of-plane deformations as buckling promoters.

Cyclic loading the samples reveals the complex role of wrinkles in graphene. Figure 3-a displays the deformation of a graphene subject to multiple compression/decompression strokes. The surface pressure follows the modulation of the accessible area of the trough (area between the barriers) in a synchronized fashion (top panel). The cyclic deformation of graphene, however, is unsynchronized as there is a finite additive delay per cycle (top vs bottom panels in Figure 3-a). Intriguingly the delay prolongs in a way that for the last cycles, increasing (decreasing) the compressive loading coincides with expanding (shrinking) graphene. The observed delayed response to loading is a manifestation of the viscoelasticity (*30*). Indeed viscoelastic materials "remember" all the loading history and the present deformation can be a response to the past loading. Note that the surface pressure (shown by + symbols in Figure 3-a top panel) modulates in synchronized manner with the accessible trough area for lipids (blue continuous line); This synchronization rules out any contribution of the lipids in our detected viscoelasticity (*31*).

Complementary to the dynamic test, viscoelasticity can be characterized by probing the gradual deformation of the material subject to a sudden loading, known as "transient creep test" (*30*). Here, we rapidly ramped up the surface pressure in equal steps of 15 $m$N/m (top panel in Figure 3-b, two steps). Obviously the graphene keeps deforming even after the stabilization of the surface pressure (bottom panel). Fitting our experimental results with the known Voigt/Kelvin model for the creep $\left(\varepsilon = \varepsilon_0 e^{-t/\tau}\right)$ (*30*, *32*) reveals characteristic relaxation times in the range of $\tau$ = 38 ± 7 s for both of the steps. We note that the dynamic test revealed signatures of the dynamic creep (*33*) as the evolution of the maxima and minima in Figure 3-a (dashed lines in the bottom panel); the observation implies an additive plastic deformation in each loading cycle.

Wrinkles in graphene exhibit considerably different configurations while progressing the compression (Supplementary Section S9). Associated with the evolution of the wrinkles, local processes such as sliding between self-contacts in curvature-localization segments and/or rotation and transition between wrinkle junctions (*34*) can delay the strain and cause viscoelasticity in graphene. Similar structural dynamics has been reported as the origin of viscoelasticity in elastomeric gels and polymers (*35–37*) and carbon nanotube ensembles (*38*, *39*). The intertwined network of wrinkles has also led to viscoelasticity in thick graphene layers (graphene paper) (*40*). Notably, the fast failure of our $Ar^+$ bombarded graphene samples does not provoke the gradual transformation of wrinkle patterns which explains the absence of viscoelasticity in those samples (Supplementary Figure S10). The assumption of the transformation of the wrinkles as the origin of the viscoelasticity is further backed-up considering that the creep tests in elastic region did not exhibit any sensible deformation delay (Supplementary Section S12).

In this context, the comparison of our observed viscoelasticity in graphene with previous reports with graphene oxide (GO) is elucidative: Indeed the amplitude of the stress response to modulating strain was used to identify transition between different phases (viscous to viscoelastic to elastic) in a LB trough (*41*, *42*). The observations are, however, considerably different both in nature and methodology: In-fact the viscoelasticity in GO film (*41*) is a result of the sliding and friction between individual GO flakes as they touch and exhibit face-to-face sliding at low surface pressures (*42*). The viscoelasticity is more the property of the floating GO ensemble, rather than the intrinsic property of a single flake. Viscoelasticity in a continuous graphene layer, however, is due to the reformation and sliding between the wrinkles, formed at high surface pressures. In terms of the methodology, we note that, the amplitude of the stress (strain) response, is unable to sign viscoelasticity, while as a delayed response is rather expected as a result of the internal friction. Our approach uniquely enables to detect the phase shift between the stress and strain in dynamic tests and to measure free relaxation in creep tests, following the protocols in classical textbooks.

Viscoelasticity in graphene is different than in conventional 3D materials (e.g. gels and polymer blocks) in the sense that i) the phenomenon is highly directional and emerges only in the presence of lateral (and not perpendicular) forces; Additionally, ii) the viscoelasticity in graphene is a result of its plastic deformation (wrinkling) at the end of an elastic regime. The peculiar thermal properties that graphene exhibits, e.g.

negative coefficient of thermal expansion and outstanding thermal conductivity, now combined with viscoelasticity will open unexplored research areas.

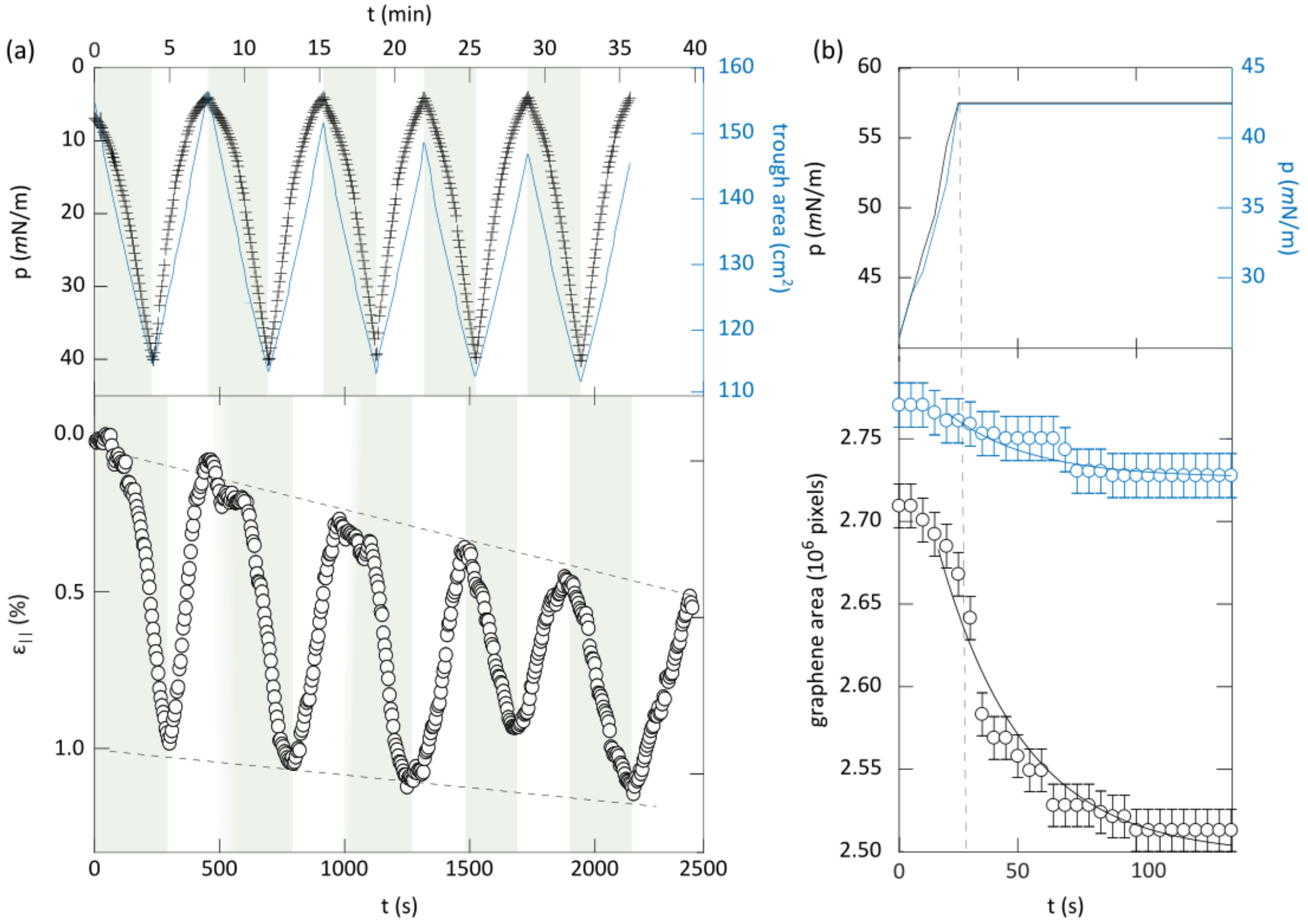

Figure 3: Viscoelasticity in graphene under compression

a) Dynamic test for viscoelasticity: graphene is subject to a cyclic loading/unloading sequence (top panel) which modulates the strain (bottom). Black data points corresponds to the surface pressure (lefts Y axis) and the continuous blue line corresponds to the available trough area (right axis). The shaded segments correspond to the loading strokes manifesting additive dephasings between stress and strain. The measurement was performed with a barrier displacement rate of 5 mm/min.

b) Transient creep test for viscoelasticity: (top panel) Graphene is rapidly loaded and stabilized in two steps of equal sizes (black and blue curves, corresponding to the left and right Y axes respectively). (bottom panel) Graphene area gradually changes in response to the loadings (the same color codes). Solid lines in the bottom panel express the best fittings with the Voigt/Kelvin model. The error bars corresponds to the uncertainty in measuring the area of graphene estimated as a band of two-pixel width at the perimeter of graphene. The rampings in surface pressure were performed with equal barrier displacement rate of 12 mm/min.

Graphene floating on water remains in its natural form and now unlocked a versatile and easily accessible platform to study the mechanics of graphene, disclosing astonishing size-dependent in-plane and flexural rigidities. Particularly the Young's modulus of centimeter-scale graphene falls two-orders of magnitude below that in micrometer sized sheets, commonly measured earlier; The finding correct the regard to graphene as the strongest material ever measured. Chemical alteration of the graphene lattice remarkably affects the mechanics of the sheet. Particularly the inclusion of vacancy defects leads to a catastrophic failure in-which a negligible surface pressure causes graphene to collapse. Hydrogenation, however, locally stabilizes graphene on water leading to an improved in-plane stiffness. Compression beyond a certain limit buckles graphene, characterized by large (~100s of nm) amplitude wrinkles which evolve upon proceeding further in compression. Most remarkably, the rheological transformation of the evolving wrinkles causes viscoelasticity, observed here for the first time. Compressing graphene floating on water, per se, provides new horizons both in science and application of graphene: piezoelectricity, surface chemistry, mechanics, material science, sensors, to name a few.

**Data availability.** The data that support the findings of this study are available from the corresponding author on request.

## S1. Methods

### Preparation of graphene samples

Graphene was grown in a CVD oven, "nanoCVD" commercially available from "Moorfield Nanotechnology". Briefly, the copper foil (Alfa Aesar, 99.999% purity, 25 µm thickness) was rapidly (in ~ three minutes) heated up to 1035 °C in the chamber and annealed under the continuous flow of hydrogen (20 sccm) for 90 minutes. Graphene grows upon the injection of methane ($CH_4$:$H_2$ ratio of 2:20) for 2 minutes; then the hot stage assembly cools down under Ar environment. The size of the hot-stage of our CVD chamber allows the growth of the samples as large as 2.5 cm × 4 cm; graphene on the backside of the copper foil was etched away upon the exposure to oxygen plasma while the front side was protected with a piece of aluminium foil, sealed at the edges with scotch tapes. We avoided using the PMMA protection as the PMMA left-over might have added uncertainties estimating the mechanical properties of graphene. The absence of the D peak in Raman spectroscopy of the front side of the randomly tested samples at this stage, confirmed the quality of the sealing. Later and by using a sharp blade, this graphene on copper sample was sectioned into pieces of ~1 cm × 1 cm to be used separately at different compression experiments.

### Compression

A piece of graphene on copper foil was placed on the commercially available Langmuir-Blodgett (LB) trough (Minitrough 2, KSV Instruments controlled by KSV Research Lab v2.01 software), pre-filled with ammonium persulfate (APS) solution (500 mM). Before the surface deposition of the lipid containing solution, the surface of the sub-phase was aspirated to minimize any contamination, indicated by the surface pressure < 0.2 mN/m after full compression. We surrounded the sample with the 1 g/L solution of 1,2-dipalmitoyl-sn-glycero-3-phosphocholine (DPPC) lipids (Avanti Polar Lipids Inc.) in $CHCl_3$/$CH_3OH$ (3:1 volumetric ratio), added dropwise at the air/water interface with a precision needle mounted on a glass syringe.

The copper foil was gradually etched away (etching time ~ 30 minutes) yielding graphene to be floating on the surface of the APS solution. We noticed a slight increase of the surface pressure (≤ 1*m*N/m) after the copper etching in different experiments. Next, two parallel hydrophilic barriers made of Delrin, slightly immersed in the liquid subphase, were moved towards each other. Depending on the specific test (normal compression, dynamic and creep viscoelasticity measurement test), the barrier displacement rates ranged from ~1 mm/min to 12 mm/min. The Wilhelmy plate, positioned close to the center of the trough, was used to measure the surface pressure. The plate was oriented parallel to the barriers, although no remarkable effect due to its orientation was detected. Indeed the isotropic surface tension is independent of the orientation (1). The LB trough was mounted on an anti-vibration table to limit the effect of external vibrations. The experiments were performed under ambient conditions with no particular controls over laboratory temperature (21 °C < T < 23 °C) and humidity level.

**Transferring graphene**

For optical microscopy, SEM and Raman characterizations, we transferred compressed graphene sheets onto SiOx/Si wafers. The process was done in the Langmuir-Schaefer mode of the set–up (2, 3) where a precisely controlled lift assembly (see Figure S1) horizontally lowers the wafer and gently place in contact with the graphene, while holding the compression. Right after the transfer, the samples were rinsed by continuous flow of ultrapure water to remove the etchant and then dried with a stream of argon.

**Manipulating the lattice of graphene**

Ar ion irradiation and hydrogenation of the graphene lattice were performed in a capacitively coupled RF plasma system operating at 40 kHz and 200 W power from Diener electronic. The base pressure of this system is below 0.02 mbar. Thorough flushing 3 to 5 times of the chamber with the corresponding gas before plasma process is necessary to ensure a high purity. The samples were placed in a vertical distance of ~3 cm away from the electrode and surrounded in a Faraday cage with gird to shield any potential bombardment from energetic ions in the plasma. The intensity of the plasma was controlled by tuning the power and pressure of the corresponding gas. The optimal parameters for controlled defect introduction are 20 W/0.9 mbar for Ar and 10 W/1.0 mbar for $H_2$, respectively. For different experiments explained in the text, different irradiation durations ranging from 30 s to 15 min were used.

**Photography**

The area of graphene during the experiments was recorded by high resolution photos captured by a Canon Eos 100D photo camera (effective pixels: 18 megapixels, maximum resolution: 5184 × 3456) equipped with a macro lens (Canon EF-S 60mm f/2.8 Macro USM). The camera was positioned perpendicular to the LB trough (inspected by a sprit level) to minimize the calculation uncertainties. In the optimized condition, we estimate deformations as small as few micrometres in each side of the square graphene to be detected by the optical set-up.

**Image processing**

We performed semi-automatic segmentation of the surface of graphene by means of image registration using *elastix* (*4*). Mathematically, the registration problems are formulated as an optimization problem to find the optimum transform parameters that is best solution to map the fixed image domain to the moving image domain. The image at time = 0 is manually segmented and a template is created from this image. The template is the moving image here and all the subsequent images are counted as a fixed image. By registration, the manual segmentation of the first image can be mapped to other images. An affine transform with the cost function of mutual information is used, which is optimized by adaptive stochastic gradient decent over 1500 iterations.

Code availability: The code used for the image processing is available from the corresponding author on request.

**Raman spectroscopy**

Raman spectroscopy was performed at the ambient conditions using WITec alpha300 R confocal Raman microscope. The setup is equipped with a dual-axis XY piezo stage providing a minimum step size of 100 nm for sample positioning. Confocal Raman spectroscopy (532 nm excitation wavelength) was used to characterize the graphene samples transferred onto wafers. In each experiment, we adjusted the accumulation time and the number of iterations to achieve a decent signal over noise ratio. The laser power was limited to < 2 mW to prevent any laser induced heating of the samples. High resolution spectra were recorded using a 100x Zeiss EC Epiplan-Neofluar DIC objective (NA = 0.9) using 600 lines/mm and 2500 lines/mm spectrometer grating.

### S2. Experiment set-up

The experiments leading to this report were performed using a commercially available, computer controlled and programmable Langmuir–Blodgett trough from KSV Instruments (model Minitrough 2). Conventionally, the trough has been used to prepare Langmuir films of different molecules and deposit mono- and multilayers onto rigid substrates. Different components of the set-up are marked in Figure S1. The trough has the dimensions of 320 mm × 75 mm × 10 mm. There is a dipping well at the centre of the trough, useful for the vertical deposition of the films with a dipping/transfer lift mechanism. In our experiments, however, we filled the well with a Teflon block and used the lift for transferring compressed graphene onto oxidized silicon wafers. A micro-electronic feedback system controls the surface pressure which is measured based on the Wilhelmy principle, using a platinum plate connected to a sensitive balance.

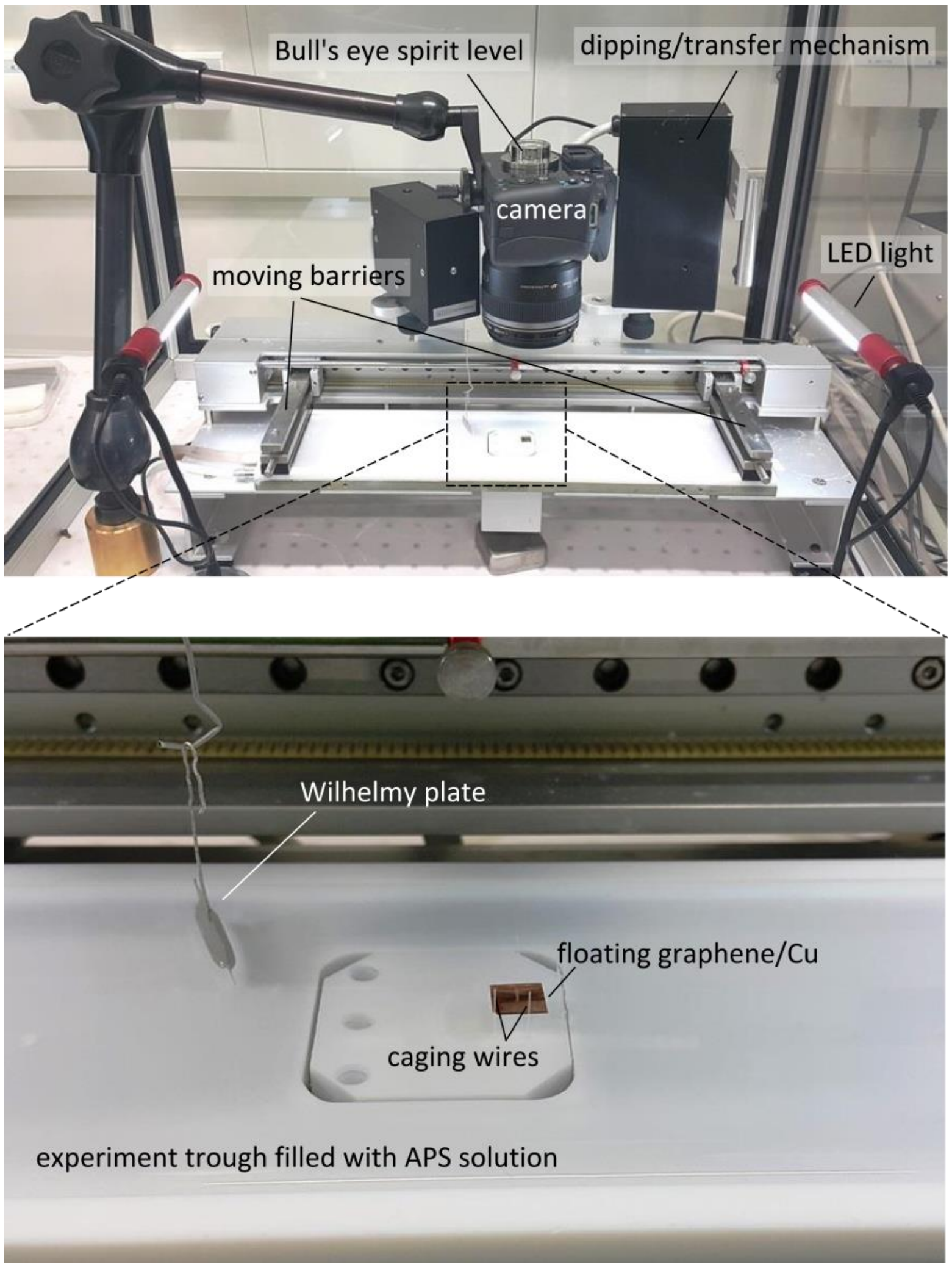

Figure S1: The Langmuir–Blodgett trough customized and used for the experiments in this report: different components are marked on the photos.

In a typical experiment, we fill the trough with 0.5 M solution of ammonium persulfate (APS) in water and float a piece of ~1 × 1 $cm^2$ of graphene grown on copper foil on the surface. The sample is surrounded and caged by few wires (with the diameter of few hundreds of micrometres) implanted in and sticking-out of the Teflon block at the bottom of the trough. The cage prevents large movements of the sample during the copper etching and later. A high resolution camera is focused perpendicularly on the sample to record its deformation during the experiments.

### S3. Rupture index as a measure of the integrity of graphene samples

The presence of microscale cracks in CVD graphene adds uncertainties to the estimated strain upon the compression of graphene. By applying a slight external compression, the size of the graphene changes rapidly (without much resistance) to fill out such open areas, leading to underestimating the stiffness of graphene. Mechanical deformation of the copper foil hosting the graphene and the vibrations of the trough, during etching the copper foil are the two important sources of the cracks. Accordingly, deliberate manipulation of the samples during the multiple preparation tasks (see the Methods) and using anti-vibration tables can minimize the crack formation.

Rupture index was previously introduced as is a measure of the density and size of the sub-micrometre cracks in graphene (*5*) which can be used to estimate the uncertainty due to the presence of the cracks in estimation of the strain. Briefly, the surface of graphene on oxidized silicon wafer is covered by PMMA layer in which a finite amount of Rhodamin B solution (4 mM in acetone) is dissolved (2:100 ratio). In a peculiar energy transfer mechanism, the conductive graphene sheet quenches the emission of the fluorophore, providing a strong contrast with the cracked areas in fluorescence microscopy images. Indeed the visibility of the cracks with fluorescence quenching method surpasses the optical and scanning electron microscopy techniques. A standardized image processing protocol further improves the contrast and convert the fluorescence microscopy images to binary arrays. Surface rupture index ($RI_s$), defined as the total number of the white pixels in an arbitrary window of 1000 pixels in the image, provides a quantitative measure for the crack coverage. The highly cracked graphene sample shown in is of $RI_s$= 81; At the early stage of the compression of such a cracked graphene piece, the external pressure would close the crack areas and result in an additional 8.1% strain. The graphene samples used in the experiments (summarized in Figure 1-c main text) were free from visible cracks; the random investigation of the samples revealed $RI_s < 1$ adding negligible error in the total strain. We note that to further compensate for this error, the early stage of the compression ($p < 10$ *m*N/m) in which the cracks get closed, was skipped in the calculation of the stiffness of graphene samples.

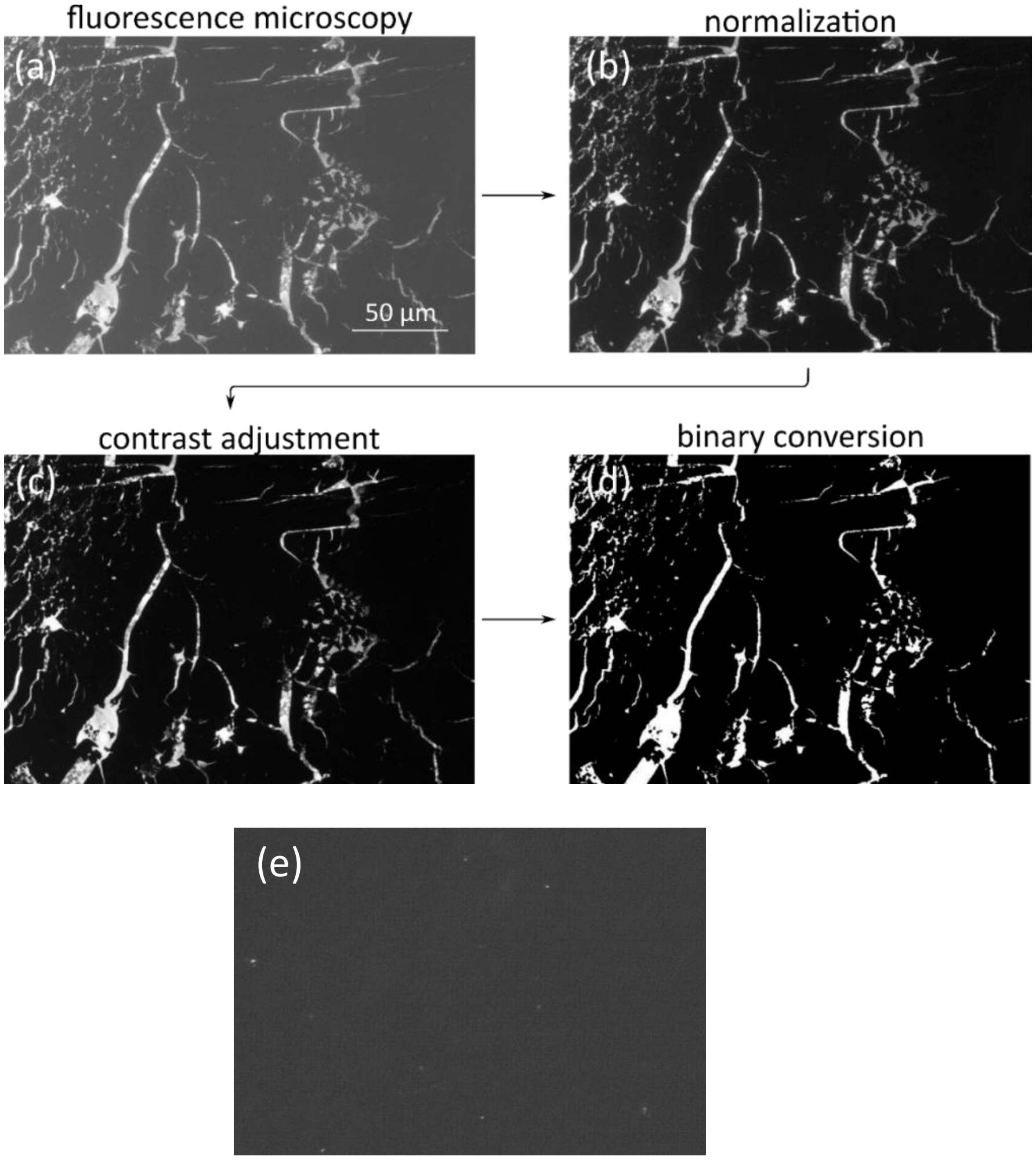

Figure S2: process flow for estimating the surface rupture index ($RI_s$):

a) The fluorescence quenching microscopy image of a heavily cracked graphene sample: the image was captured with 20X objective in a fluorescent microscope, while the surface of graphene was pre-coated with a suitable amount of PMMA mixed with 4mM of Rhodamine B solution (2:100 ratio).

b) The image is normalized to eliminate the effect of any intensity variations from image to image.

c) The contrast improves by saturating 1% of the data with the lowest and highest intensities.

d) A median threshold value is used to convert the gray-scale image into the binary; The $RI_s$ is estimated as the averaged sum of the white pixels in arbitrary windows of 1000 pixels in this image. The process is repeated for several (at least eight) images captured at different spots of the same sample to provide an average $RI_s$.

e) Representative FQM of a crack-free graphene sample used in the experiments of this manuscript.

The scale bar in panel (a) holds for all the figures.

### S4. Size-dependent in-plane elastic properties of graphene

Monte Carlo simulations demonstrated that the elastic properties of graphene in the presence of anharmonic coupling the in-plane and out-of-plane displacements are highly size dependent (*6*). Particularly the equilibrium bulk modulus of graphene vanishes with the graphene size as $L^{-0.323}$ following the relation:

$$k = \frac{193.6 - 9.1 \times 10^{-2} L^2 + \frac{457.6 \left(\frac{L}{14.14}\right)^4}{L^{0.325}}}{1 + \left(\frac{L}{14.14}\right)^4} \qquad \text{(s1)}$$

Here L is the length (in Å) and k is the bulk modulus (in N/m) of graphene sample. Bottom inset Figure 1-c superimposes the theoretical and experimental results. Obviously, the size dependency of k can largely explain the considerably low (with respect to earlier reports (*7–11*)) in-plane stiffness we measured. The median k of our samples, however is slightly above the theory expectation. The suppression of the anharmonicity due to the defects in CVD graphene (as reported previously (*11*)) may account for this discrepancy.

### S5. Manipulation of the graphene lattice

Inclusion of the crystalline defects upon exposing the graphene samples to argon and hydrogen plasmas modify the mechanics of graphene. The defect density $n_D$ can be estimated using the relative intensities of the Raman D and G peaks (*12*):

$$n_D\ [cm^{-2}] = (7.3 \pm 2.2) \times 10^9 E_L^4 (I_D/I_G)$$

In this relation $E_L = 2.33$ eV is the excitation energy. Figure S3 shows the Raman spectra and defect characterization for different samples. Note that the intercept of the polynomial fitting of the time dependent defect density gives $n_0 = 9 \times 10^{10}$ cm$^{-2}$ for untreated graphene samples.

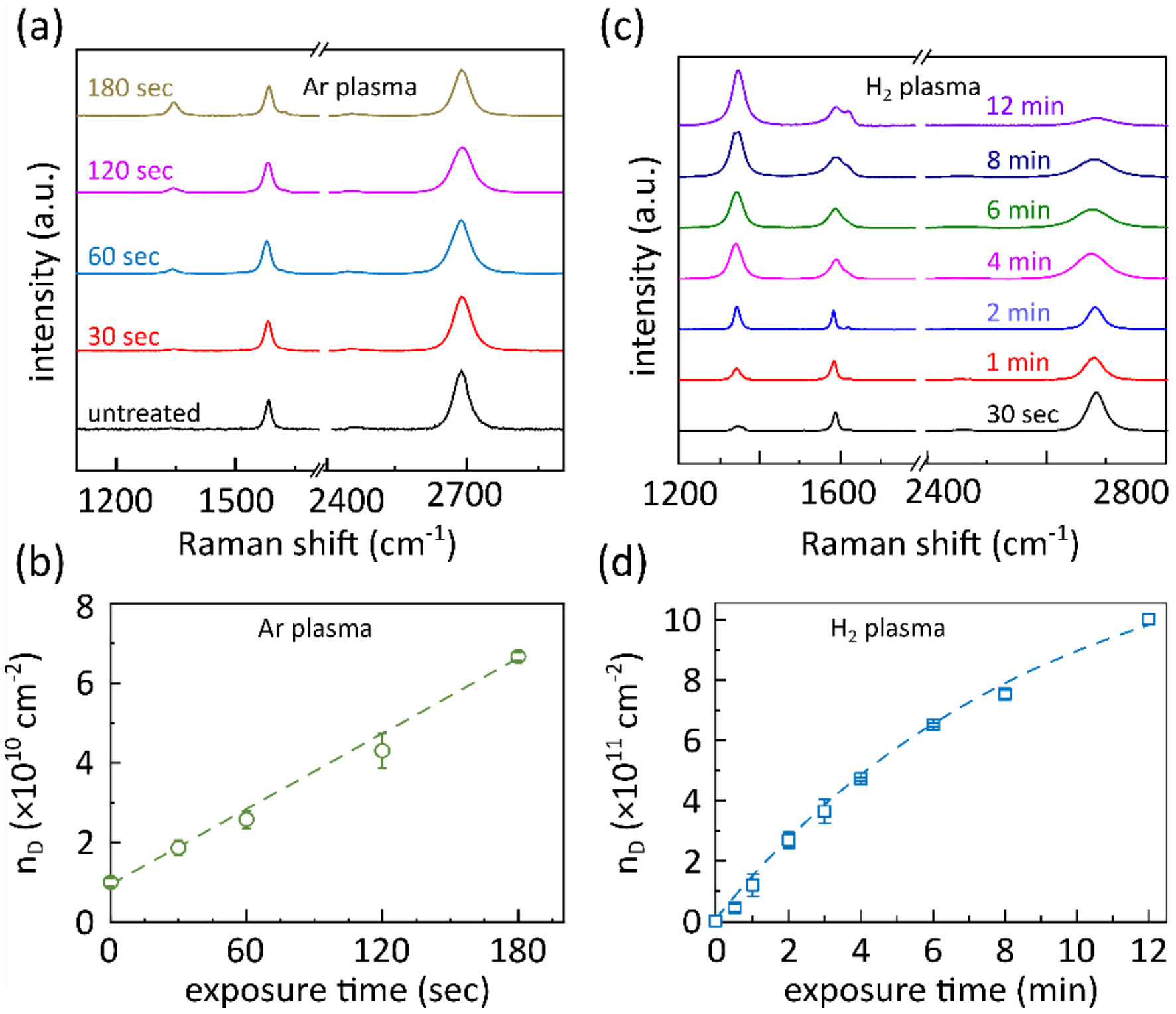

Figure S3: Raman characterization of functionalized graphene samples
a) Typical Raman spectra recorded on graphene samples exposed to Ar plasma for different durations:
b) Density of the vacancy defects ($n_D$) calculated for different samples in a.
c) Typical Raman spectra recorded on graphene samples exposed to $H_2$ plasma for different durations
d) Density of the $sp^3$ hybridization defects calculated for different samples in c
All the Raman spectra were recorded on graphene/silica wafers, using an excitation source of $\lambda = 532$ nm.

Indeed the Raman measurements show that the density of the defects for $Ar^+$ plasma bombardment and hydrogenation reach maximum of $7 \times 10^{10}$ $cm^{-2}$ and $1 \times 10^{12}$ $cm^{-2}$ respectively. Although the density of such defects seems rather negligible considering the density of the carbon atoms in the graphene lattice (~$3 \times 10^{15}$ $cm^{-2}$), the effect on the mechanics of graphene is rather pronounced (see Figure 1-c and d of the main text).

Energetic $Ar^+$ ions during the plasma bombardment of graphene are capable to effectively knock-out carbon atoms from the lattice and generate vacancy defects of atomic sizes (*13*). Hydrogen radicals, on the other hand, are of limited momentum (compared to Ar ions); they are chemically active and form covalent bonds to carbon atoms, altering their hybridization state from $sp^2$ to $sp^3$ (*14*, *15*). The plasma treatment is performed under precise control of the gas composition: we flushed several times the reaction chamber prior to the plasma treatment to eliminate the residual gases and prevent any unknown/unfavorable functionalization.

## S6. Mechanics of graphene with defected crystalline lattice

$Sp^3$ hybridization of carbon orbitals − as a result of reaction with various chemical components − and vacancy defects − generated by shooting-out carbon atoms in momentum transfer processes − are the two major defect types commonly observed in graphene lattice. We characterized the mechanics of graphene in the presence of such defects. Hydrogenation of graphene and argon bombardment of the lattice in a plasma chamber respectively achieved $sp^3$ hybridization and vacancy defects. Technical details are covered in the Methods section.

Figure S4-a details the stress-strain plots of graphene samples subject to $Ar^+$ bombardments of up to 180 s. Reduction of the critical pressure at the cross-over between linear (elastic) and non-linear (plastic) deformations upon increasing the vacancy dose is obvious in the plots. Even after 180 s of plasma treatment, the relative concentration of the vacancies ($\sim 6 \times 10^{10}$ cm$^{-2}$) to the concentration of the carbon atoms in graphene lattice ($3.82 \times 10^{15}$ cm$^{-2}$) is negligible; hence the treatment is not expected to (and does not) affect the rigidity as it is originated from the degree of order and the strength of the bonds between the atoms (*16*).

Hydrogenation of graphene, in contrary, increases the $p_{cr}$. We attribute the effect to the increased interaction between hydrogenated graphene lattice and water substrate (e.g. hydrogen bounds) which improves the in-plane Young's modulus upon partial screening of long wavelength flexural modes. It seems increasing the dose of the $sp^3$ defects ultimately (and gradually) turns to weaken the lattice (*11*).

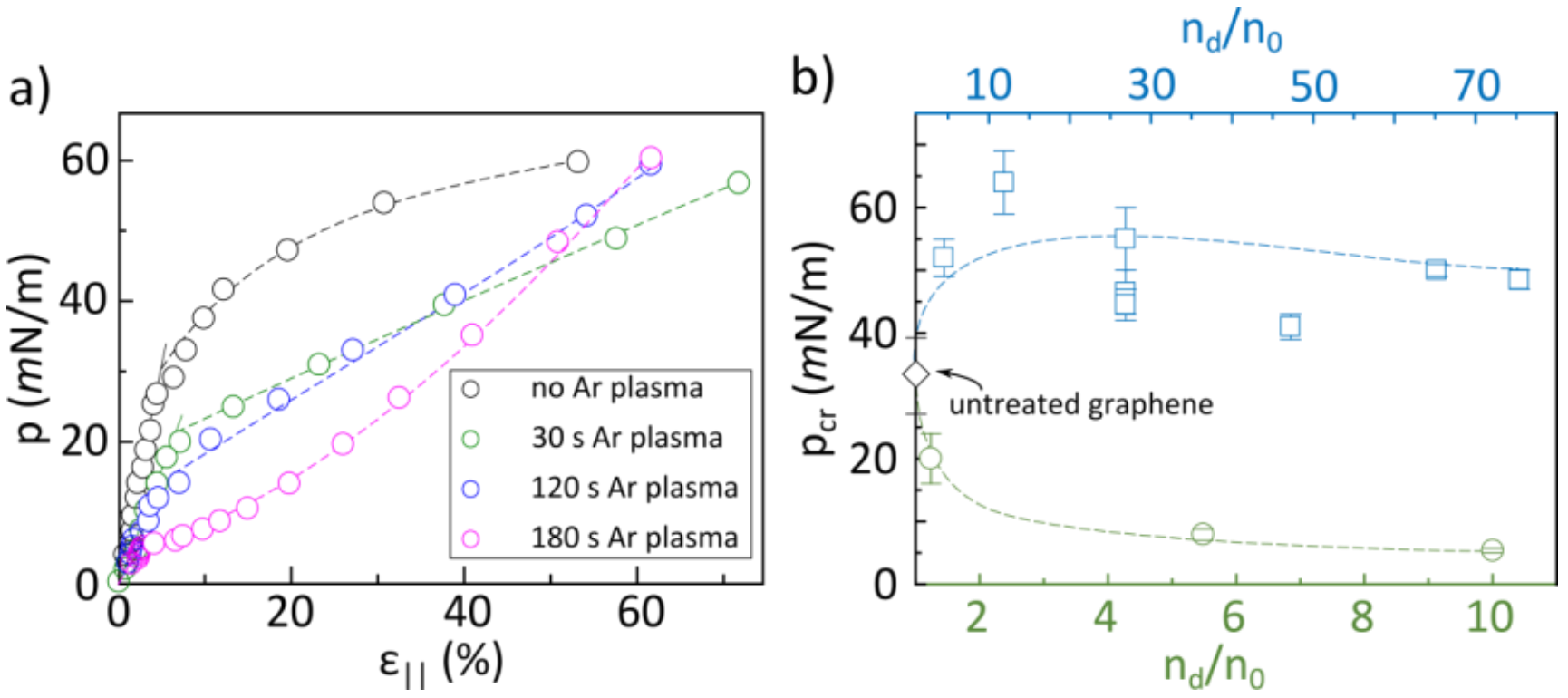

Figure S4: Mechanics of the graphene subject to $sp^3$ hybridization and vacancy defects
a) Stress-strain plots of graphene samples subject to Ar ion bombardment of different durations
b) Critical pressure of hydrogenated and Ar ion bombarded graphene samples: the density of the defects ($n_d$) achieved by the Raman spectroscopy (Figure S3) is normalized by the defect density of untreated graphene ($n_0 \sim 9 \times 10^9$ $cm^{-2}$). The $p_{cr}$ of untreated graphene is the median of all the measured samples detailed in Figure 1 of the main text.
All the dashed lines are guides to the eyes.

## S7. Graphene buckling: blistering or wrinkling

In the process of buckling, thin films delaminate from the surface of *rigid* substrates and form blisters. Similar scenario may happen on the surface of liquids provided that the film/substrate adhesion is small enough; Otherwise the liquid substrate may deform together with the film to form wrinkles. More precisely, the competition between the work of a hydrostatic pressure of the liquid (following the deformation of the film) and surface energy cost to delaminate the film (assuming that the film remains wet after the delamination) determines which buckling mechanism is in action. Wagner et al studied the compression driven bulking of an elastic film floating on a liquid surface (*17*); They introduced a dimensionless parameter $\Gamma/\delta^2$ ($\Gamma \equiv \gamma\lambda/D$ and $\delta \equiv \rho_s t/\rho_l L$, the parameters are explained in Table S1) which sets the critical blister size at the unset of the blistering. Particularly for $\Gamma/\delta^2 >> 1$ (our case), their model predicts a blistering size of ~ 10 cm, larger than the sample size implying that blistering is impractical for our samples.

Table S1: parameters involved in calculating the critical blister size, based on the model proposed by Ref (*17*)

| **L [m]** | **$1 \times 10^{-2}$** | **initial length of the sample** |
|---|---|---|
| **t [m]** | $3 \times 10^{-10}$ | thickness of graphene |
| **$\rho_l$ [kg/m$^3$]** | $1 \times 10^3$ | density of water |
| **$\rho_s$ [kg/m$^3$]** | $2.2 \times 10^3$ | density of graphene |
| **D [N.m]** | $1 \times 10^{-12}$ | flexural rigidity of graphene |
| **g [m/t²]** | 9.8 | gravity acceleration |
| **γ [N/m]** | $7.3 \times 10^{-2}$ | surface tension of water |
| **λ [m]** | $1 \times 10^{-4}$ | intrinsic wavelength of wrinkles = $(D/\rho_l g)^{1/4}$ |

**S8. Molecular dynamics simulations**

We simulated the dynamical response of a monolayer graphene subject to compression by means of large scale atomic/molecular massively parallel simulator (LAMMPS) (*18*). Adaptive intermolecular reactive empirical bond order (AIREBO) potential (*19*) was chosen to describe interatomic interactions between C-C and C-H pairs. Water molecules have been modelled by the four-point TIP4P potential (*20*) with the H-O bond length of 0.9572 Å and the H-O-H bond angle of 104.52°. The interaction between atoms accounts for both Coulomb and Van der Waals contributions. To describe the latter, we employed a 6-12 Lennard-Jones (LJ) potential, $4\varepsilon_{ij}\left[(\sigma_{ij}/r)^{12} - (\sigma_{ij}/r)^{6}\right]$, with parameters $\varepsilon_{OO}$ = 0.1627 kcal/mol and $\sigma_{OO}$ = 3.1643 Å. The long-range electrostatic interactions were calculated using the particle-particle particle-mesh (PPPM) method. Van der Waals interactions between water molecules and hydrogenated graphene films were characterized by LJ parameters $\varepsilon_{CO}$ = 0.0927 kcal/mol and $\sigma_{CO}$ = 3.2830 Å (*21*, *22*), and $\varepsilon_{OH}$ = 0.0749 kcal/mol and $\sigma_{OH}$ = 2.9071 Å, where the latter two parameters were obtained from the Lorentz-Berthelot combining rules.

The modeled untreated graphene consist of 3936 carbon atoms spanning roughly square area of $10\times10$ nm². Addition of a single hydrogen atom to this model provides a hydrogen density of $10^{12}$ cm$^{-2}$, corresponding to the experimentally measured density for hydrogenated graphene. A relatively thick layer of water molecules with height $h_w$ = 16.7 Å encloses the bottom subspace, from which the lowest water molecules have been kept rigid. Periodic boundary conditions were imposed along the two orthogonal directions parallel to the graphene. Initial equilibration run was performed under isobaric and isothermal conditions (at room temperature and zero pressure) for 10 ps. The compression process starts by applying a homogeneous biaxial strain with constant rate 0.04 $\%ps^{-1}$ within NVT simulations. In all our molecular dynamics simulations, a time step of 1 fs has been used for the velocity-Verlet integrator.

In agreement with the analytical results (Supplementary Section S7), the molecular dynamics simulations confirm the absence of blistering in our system (Figure S5-a). Indeed the water molecules follow the undulations of graphene throughout the compression process and no delamination is detectable. Hydrogenation of graphene, on the other hand, locally stabilizes the sheet around the hydrogenated site, implying locally stronger interaction with underlying water molecules (Figure S5-b). The interaction significantly affects the stiffness of graphene (Figure S5-c), resulting in higher bulk modulus than for graphene by a factor of 1.35.

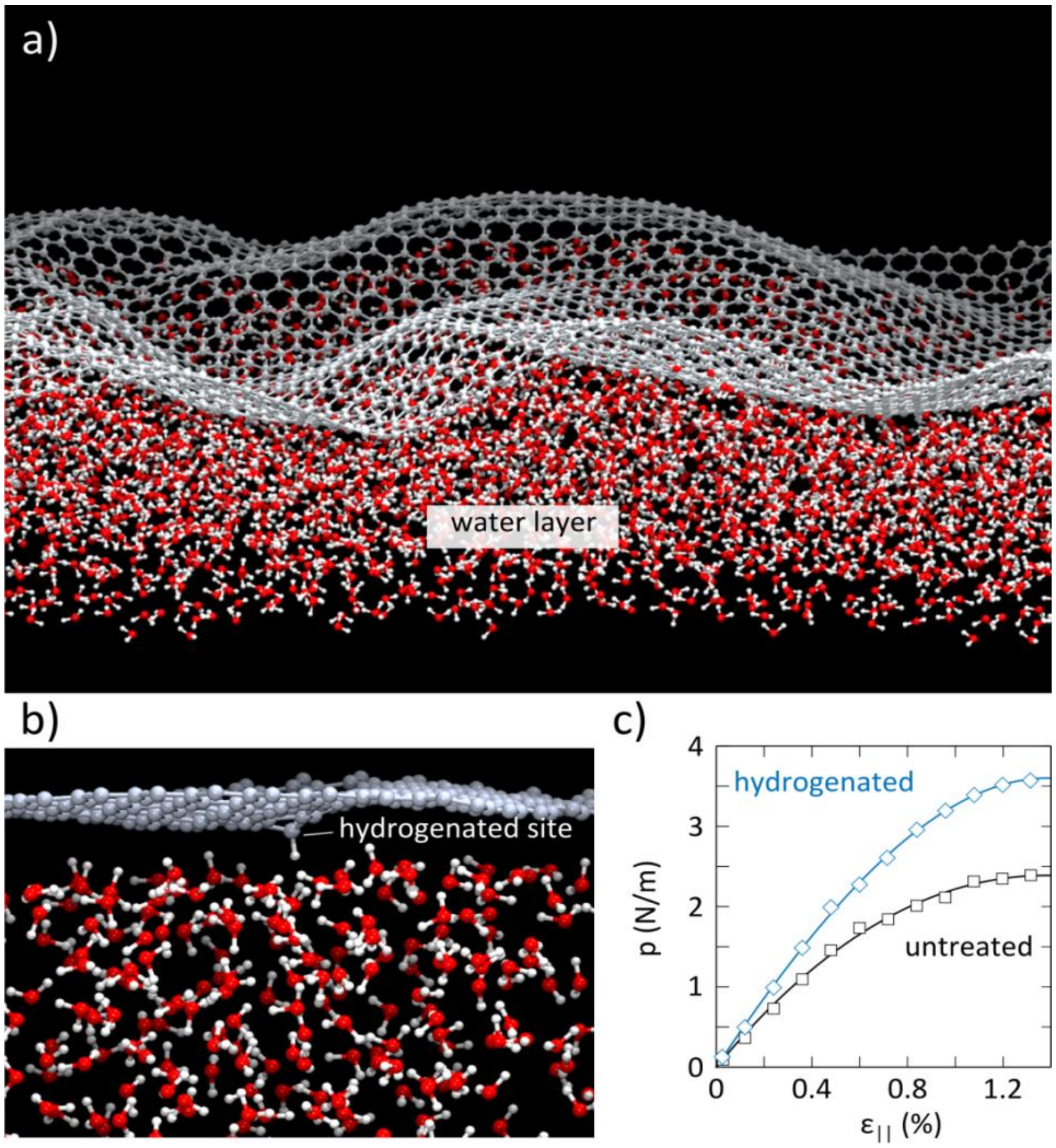

Figure S5: Molecular dynamics simulation of graphene/water interplay
a) Heavily wrinkled graphene on the surface of water: water molecules follow the undulations of graphene and no delamination is observable. The snapshot corresponds to $\varepsilon > \varepsilon_{cr}$ region.
b) Hydrogenated graphene on the surface of water: interaction between the hydrogenated site and water molecules locally stabilizes the corresponding carbon atom. The snapshot corresponds to $\varepsilon < \varepsilon_{cr}$ region.
c) Evolution of the stress-stain diagram as a result of the hydrogenation: Hydrogenation improves the Young's modulus by 1.35 times. The concentration of hydrogenated sites ($10^{12}$ $cm^{-2}$) in the simulations is similar to the experimentally measured values.

## S9. Morphological study of compressed graphene

Transferring strained graphene onto oxidized silicon wafers enables to characterize the morphology. Here at the target surface pressure and by using the dipping/transferring mechanism (Figure S1) we gently put the wafer in contact and let the graphene to transfer onto the wafer. Wrinkles appear in different patterns at different levels of strain (**Error! Reference source not found.**). Particularly while the sample compressed in elastic region $\varepsilon < \varepsilon_{cr}$ is free from visible wrinkles, buckling of graphene close to and above the $\varepsilon_{cr}$ forms numerous wrinkles. The wrinkles are initially orthogonal to graphene borders. The pattern considerably changes at a slightly higher strain ($\varepsilon > \varepsilon_{cr}$); here major wrinkles enclose flat lands and form polygonal cells. The lengths of the wrinkles may reach 100 µm and meet each other in X and Y shaped junctions. Besides a secondary pattern of wrinkles with smaller lengths are generating in-between the cells. Well above the critical strain (surface pressure of ~60 *m*N/m) a uniform landscape of short wrinkles cover the whole surface of graphene.

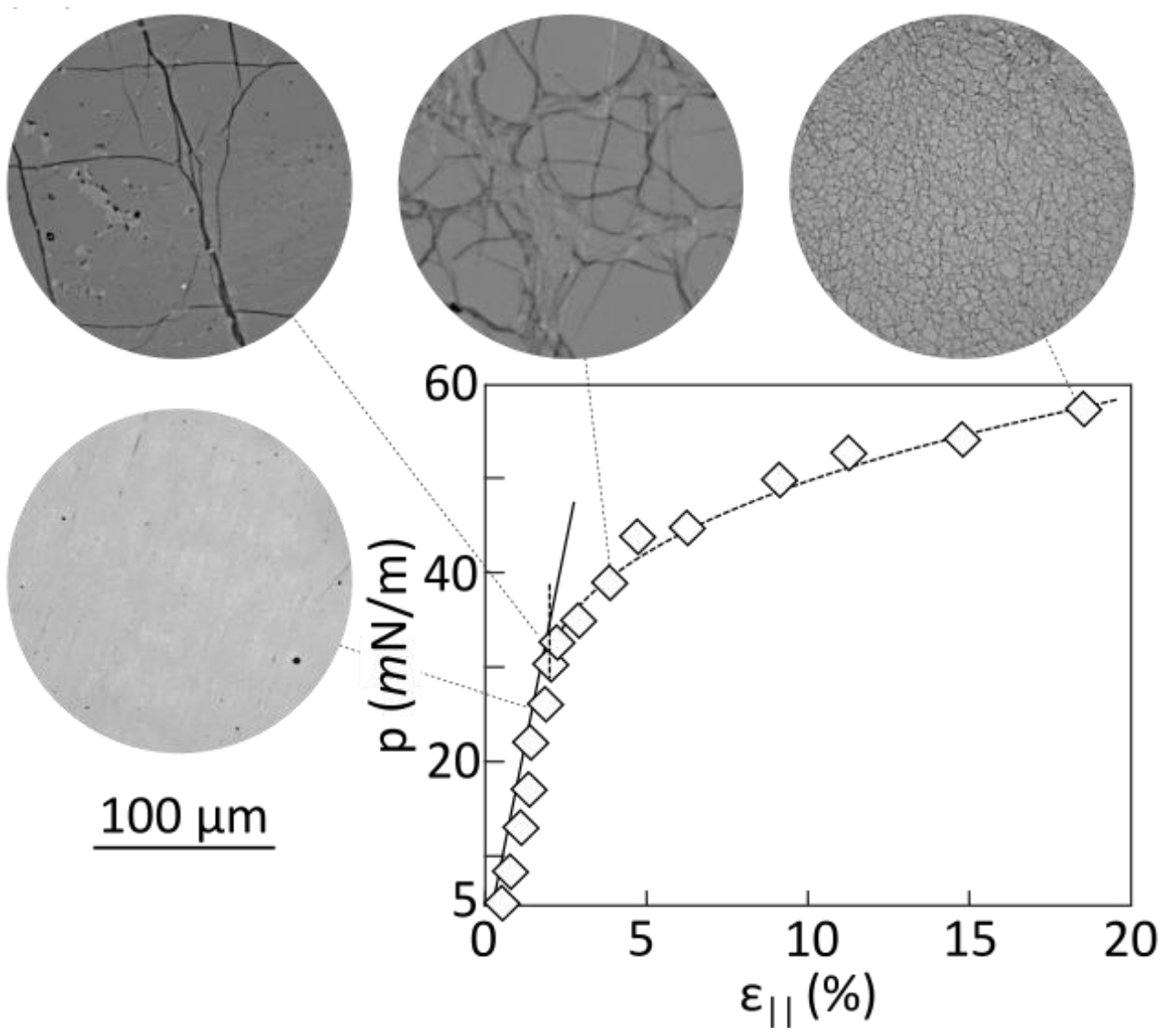

Figure S6: Morphology of graphene, strained at different levels and transferred onto oxidized silicon wafer

## S10. Catastrophic failure of Ar bombarded graphene samples

The behaviour of the untreated and hydrogenated graphene samples matches the classical picture of buckling (*23*) where the buckling strain is inversely proportional to the stiffness. The failure of the $Ar^+$ bombarded samples, however, does not follow this trend (Figure 1-d). Here the critical pressure requires to initiate permanent deformations drops considerably by the inclusion of a slight vacancy dose (Figure S4). Post-mortem analysis of $Ar^+$ bombarded samples does not show a regular pattern detected in untreated and hydrogenated graphene (compare Figure S7-a, b and **Error! Reference source not found.** for $\varepsilon >> \varepsilon_{cr}$). Particularly two different levels of folding patterns featuring dilute and dense folds − randomly distributed all over the sample − are observable. Timewise, the deformation of the $Ar^+$ bombarded samples is synchronized with the external pressure (Figure S7-d). This behaviour is distinct from untreated graphene samples where a slow transition between the wrinkle patterns led to a delayed deformation (Figure 2 of the main text). All the observations highlights that a catastrophic failure is in play now.

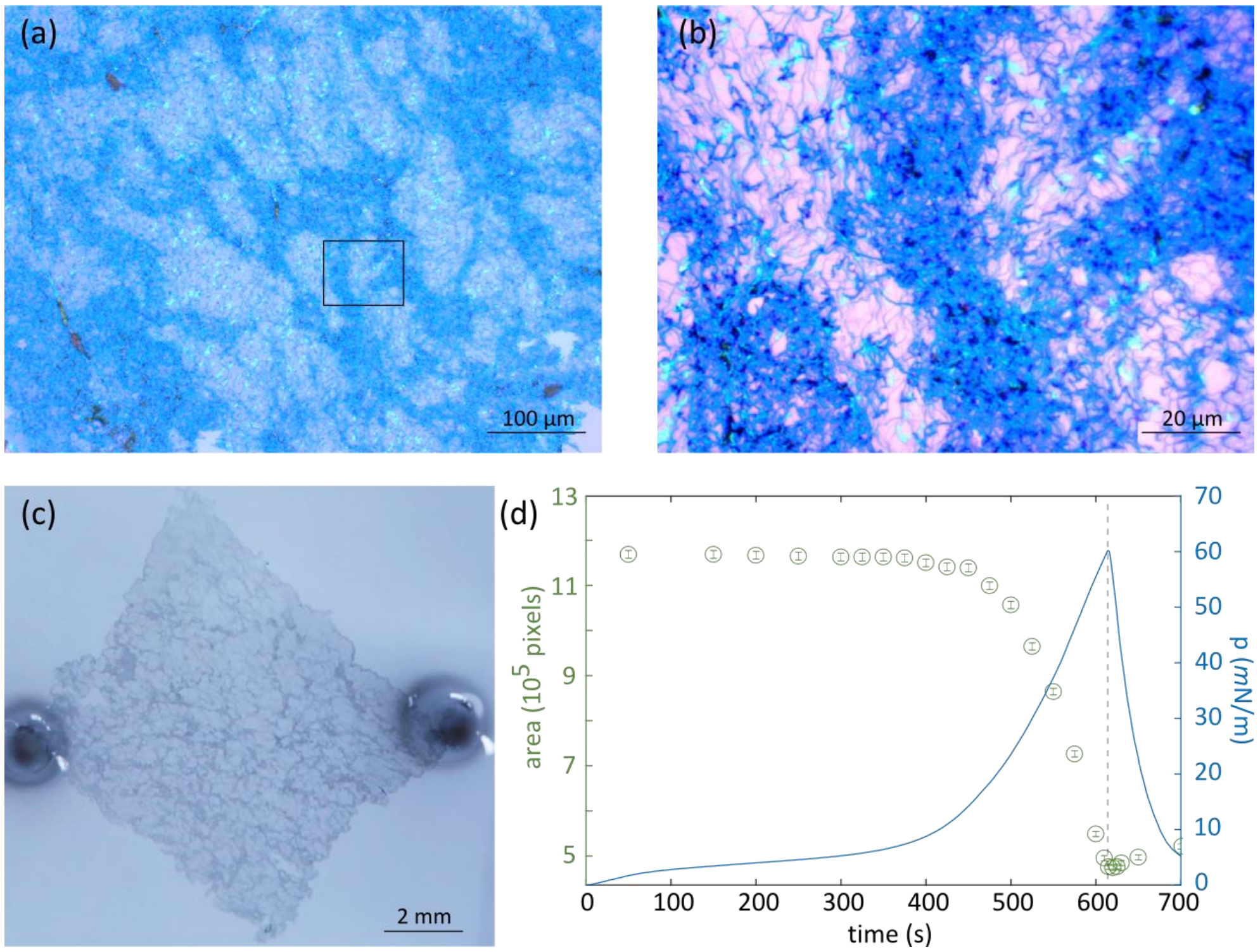

Figure S7: Characterization of the failure of $Ar^+$ bombarded graphene samples
a) Optical micrograph of an Ar+ bombarded sample (120 s of plasma) transferred on to a SiOx/Si wafer, after being compressed up to 60 $m$N/m
b) Zoomed-in view to the small window marked in a.
c) Photograph captured at the end of the compression course of the sample in a, floating on the aqueous surface; Two of the wires forming a cage for graphene are observable at the left and right sides of the graphene.
d) Estimated area of the graphene sample in c as a function of time and surface pressure

## S11. Flexural rigidity of graphene

Flexural rigidity of graphene governs the onset of buckling. Particularly with Euler's buckling formulation, the critical strain of an square sample reads (*23*):

$$\varepsilon_{|cr} = \frac{\pi^2 D}{a^2 E_{2D}}\left(m + \frac{1}{m}\right)^2 \quad \text{(a)}$$

Here $a$ is the length (= width), $D$ is the flexural rigidity, $E_{2D}$ is the two-dimensional Young's modulus and $m$ is the number of half waves created upon buckling of the sample. Accordingly the fitting $\varepsilon_{|cr} = \frac{\alpha}{(E_{2D}/\beta)}$ ($\alpha$ = 0.027 N/m, $\beta$ = 1 for untreated samples) in Figure 1d reveals $D$:

$$D = \frac{\alpha a^2}{\pi^2\left(m + \frac{1}{m}\right)^2} \qquad \text{(b)}$$

The graphene sample strained up to the onset of buckling ($\varepsilon \sim \varepsilon_{cr}$, **Error! Reference source not found.**) features numerous horizontal and vertical undulations, separated in $\lambda/2 \sim 50$ μm range providing an estimation for $m$: $m = a/(\lambda/2) \sim 200$. Accordingly the equation above suggests $D$ in the order of $10^{-12}$ N/m. Numerous half waves observed in our sample is due to the interaction of the graphene with liquid substrate which stabilizes the sheet at high energy levels. Notably in the absence of water, graphene samples would form only one large buckle ($m$ = 1). In this case equation (a) with our estimated flexural rigidity predicts a critical buckling force of $10^{-10}$ N order which is the same order of the weight of our graphene piece. In other words vertically standing graphene walls would be highly unstable and may collapse by its own weight.

## S12. Transient creep test

Transient creep test is an standard test to characterize viscoelastic materials (*24*). Here stress is sharply applied to the material and the strain response is recorded. Viscoelastic materials display gradual deformations; the longer the settling time, the stronger the viscos character. The deformation of pure elastic materials however is sharp and follows the same timeline of the applied stress. Figure S8 displays the performance of a graphene sample in this test. The surface pressure has raised in a stepwise manner from ~ 0 mN/m in equal steps of ~ 15 mN/m. The deformation of graphene in the two first steps (where graphene is still in elastic region) does not exhibit any sensible delay. In plastic deformation region (the two last steps), the deformation is gradual manifesting a remarkable viscoelastic property.

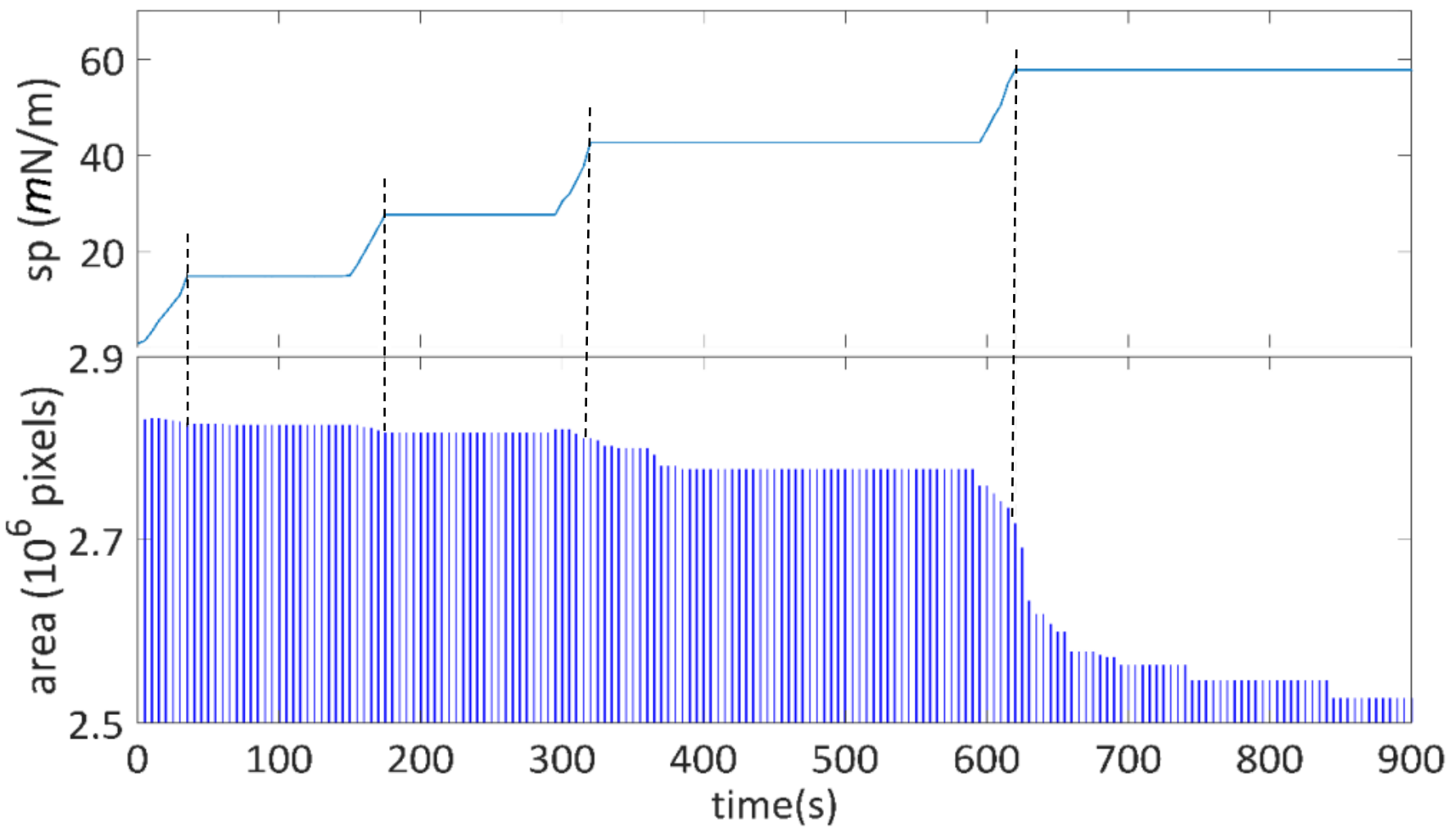

Figure S8: Transient creep test of a graphene sample
The surface pressure (top panel) is raised in stepwise manner in equal steps of ~15 mN/m while the area of graphene is recorded. The error bars in bottom panel have been emitted to improve the visibility. Figure 2 of the main text replots the last two steps of this graph.

## S13. π-A relation of the lipid/graphene film

Figure S9 demonstrate a typical surface pressure-area compression isotherm of a DPPC lipid monolayer on an aqueous solution of ammonium persulfate (500 mM) and the subsequent compression/decompression curves (corresponding to the dynamic viscoelastic test on Figure 3a of the main text). The modulating isotherm indicates a reversible compression of the lipid layer. The isotherm features a drift in the area upon sequential compression/decompression cycles (see blue circle markers in figure S9); the drift is progressive and much larger than the area of the graphene (1 $cm^2$) which can be attributed to the loss of the lipids during the experiments (collapsing into sub phase to escaping from the barriers). Indeed by means of the IR-ERS spectra of graphene at different surface pressures, we already demonstrated that the lipids do not collapse onto and/or below the graphene film (*25*). As long as the compression in the lipid layer is uniform (guaranteed by minimized compression speed) and the pressure applied to graphene remains the same as is measured by the Wilhelmy plate, the drift in isotherm is not of any experimental concern. Inset compares the single compression experiments on APS solution and on water (respectively with and without graphene). The transition between different phases are rather smooth on APS.

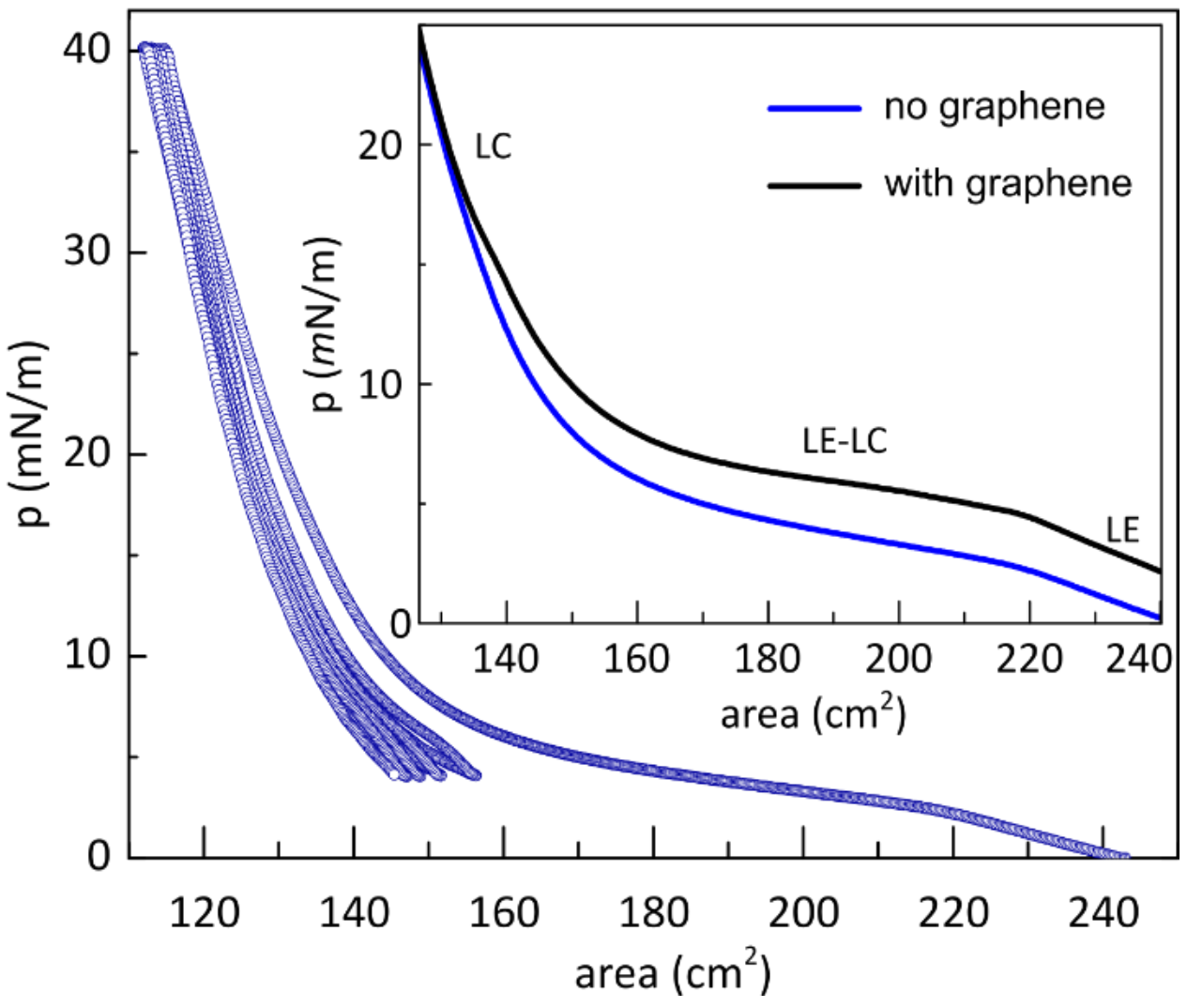

Figure S9: Typical π-A isotherm of DPPC lipids on an aqueous solution of 500mM of ammonium persulfate (APS) upon sequential compression and decompression; the data corresponds to the dynamic viscoelasticity test in the main text (Figure 3a). The inset compares the isotherm with and without the presence of graphene, respectively on APS solution (0.5 M) and on water sub phase. Different phases of lipid compression (LE: liquid expand, LC:L liquid condensed) are marked.